\documentclass[aps,prl,reprint,superscriptaddress]{revtex4-2}

\usepackage{amsmath,amssymb}
\usepackage{graphicx}
\usepackage[colorlinks=true, allcolors=blue]{hyperref}
\usepackage{upgreek}
\usepackage{orcidlink}

\begin{document}

\title{Multiple propagating modes in random polycrystals}

\author{Ming Huang\,\orcidlink{0000-0002-5868-4244}}
\email{ming.huang@sussex.ac.uk}
\affiliation{Department of Engineering and Design, University of Sussex, Brighton BN1 9RH, United Kingdom}
\affiliation{Department of Mechanical Engineering, Imperial College London, London SW7 2AZ, United Kingdom}
\author{Stanislav I. Rokhlin\,\orcidlink{0009-0002-5791-1430}}
\affiliation{Department of Materials Science and Engineering, The Ohio State University, 1248 Arthur E. Adams Drive, Columbus, Ohio 43221, USA}
\author{Michael J. S. Lowe}
\affiliation{Department of Mechanical Engineering, Imperial College London, London SW7 2AZ, United Kingdom}

\begin{abstract}
Multiple-scattering theory predicts that the coherent elastic wave in a statistically isotropic polycrystal is not a single wave. Its dispersion equation admits multiple roots of a given polarization, but these extra modes have never been observed. Here, using grain-scale finite-element simulations of three-dimensional polycrystals, we resolve the coherent field in the frequency--wavenumber plane and directly observe the predicted multiplicity for both transverse and longitudinal excitations: a single mode at low frequencies gives way to two in the geometric regime as the wavelength approaches the grain correlation length. The measured modes follow the roots of the dispersion equation, which always carries a third root whose attenuation exceeds that of the others. It is this attenuation hierarchy, not the root count, that dictates how many modes propagate and can be observed. Coherent wave propagation in random media is thus inherently multimodal in the geometric regime, which we reach here for the first time in polycrystals.
\end{abstract}

\maketitle

An elastic wave traversing a random medium loses energy from its coherent, ensemble-averaged component to the scattered field, and the surviving coherent wave is conventionally described by a single complex effective wavenumber per polarization~\cite{Ishimaru1978,Sheng2006,Akkermans2007}. Yet a statistically homogeneous random medium can carry more than one wave of a given polarization. Biot predicted this first for fluid-saturated porous solids, whose coupled solid--fluid motion supports fast and slow compressional waves~\cite{Biot1956a,Biot1956b}, confirmed ultrasonically later by Plona~\cite{Plona1980}. Brillouin-scattering measurements on colloidal suspensions then revealed two longitudinal modes once the wavelength became comparable to the sphere size~\cite{Liu1990}. These were explained as quasimodes~\cite{Jing1991}, and were later mapped in detail by pulsed ultrasonics~\cite{Page1996,Cowan2011}. Most recently, for random particulate composites, it has been proved that the ensemble-averaged field is a superposition of infinitely many effective waves rather than one wave~\cite{Gower2019a,Gower2019b}, and an independent variational treatment found that a layer of random composite supports exactly two mean plane waves~\cite{Willis2024}. In all of these systems, however, the multiplicity is tied to the presence of two constituents.

Single-phase polycrystals pose the question in its purest form. There is no second constituent, and the only disorder is the random orientation of elastically anisotropic grains. For scalar waves in continuous random media, Frisch~\cite{Frisch1968}, following Tatarskii and Gertsenshtein~\cite{Tatarskii1963}, gave the mean Green's function as two terms with distinct wavenumbers, whose attenuations and amplitudes become comparable at high frequency (see also~\cite{Rytov1989}), and Calvet and Margerin~\cite{Calvet2012} showed that the second root is demanded by causality, because a single mode with the asymptotic velocity and attenuation of the geometric limit cannot satisfy the Kramers--Kronig relations. For elastic waves, the far-field approximation of Rokhlin \textit{et al.}~\cite{Rokhlin2015}, like the second-order theory of Stanke and Kino~\cite{Stanke1984}, yields two longitudinal and transverse roots. In polycrystals, however, this multiplicity has remained purely theoretical: no additional mode has ever been observed, and it has not been known whether the extra roots are physical waves or artifacts of the approximations that produce them.

In this Letter we supply that observation for statistically isotropic polycrystals. Grain-scale finite-element simulations, which solve the elastodynamic equations directly and invoke no scattering theory, with the coherent field examined in the frequency--wavenumber plane, resolve two transverse modes and two longitudinal modes, emerging when the wavelength becomes comparable to the grain correlation length. The simulations also show the mechanism directly: the injected plane wave is extinguished within about one correlation length, and the coherent field beyond is rebuilt by forward multiple scattering, whose distinct longitudinal and transverse channels support distinct modes. Comparison with theory then completes the picture. The dispersion equation admits three roots at all frequencies, but only in the short-wavelength geometric regime do two of them attain comparable attenuation while retaining distinct phase velocities, becoming separately observable. The third root attenuates faster throughout and cannot be observed at any frequency. This means that the number of propagating modes is set not by the number of roots but by their attenuation hierarchy. We demonstrate these findings in two representative cubic materials of common and extreme anisotropy, and the material choice only shifts the frequency at which the geometric regime occurs.


\begin{figure*}[!ht]
  \includegraphics[width=0.72\textwidth]{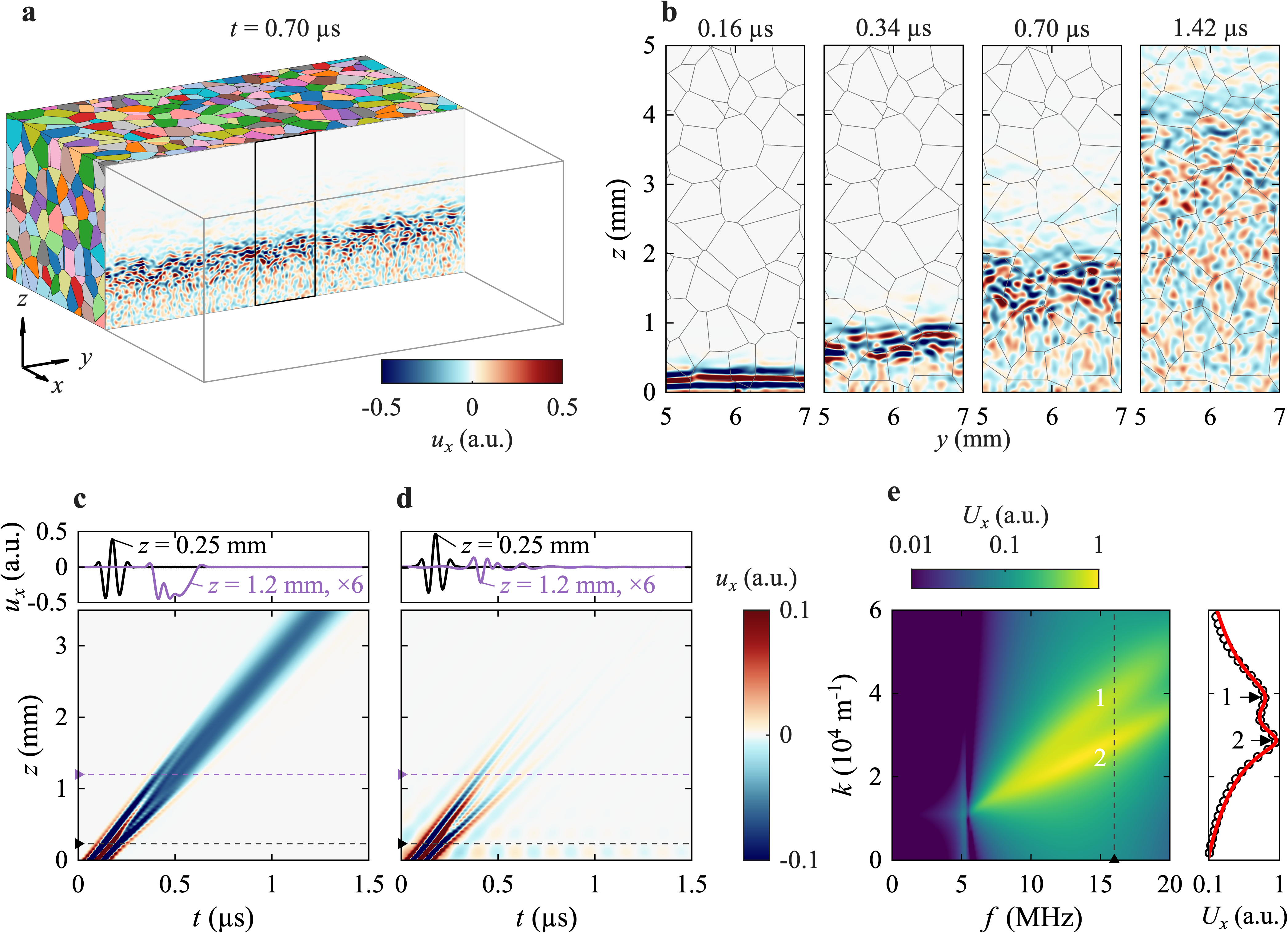}
  \centering
  \caption{Finite-element simulation of a plane transverse wave pulse in statistically isotropic polycrystalline iron; the pulse is excited on the $z=0$ plane. (a)~Computational domain cut open at $x=6$~mm, with the grains colored arbitrarily and the removed half outlined in gray; color on the cut face is the displacement $u_x$ at $t=0.70~\upmu$s in a single realization at 16 MHz excitation ($2k_\mathrm{0T}a_\mathrm{C}=14.37$), and the black rectangle marks the window of (b). (b)~$u_x$ on the same plane at the four times shown, overlaid with the grain boundaries on that plane. (c)~Coherent field $u_x(t,z)$, averaged over the $x$--$y$ cross section at each depth $z$ and over 20 realizations; the strip above shows the waveforms at $z=0.25~\mathrm{mm}\simeq a_\mathrm{C}$ and at $z=1.2~\mathrm{mm}\simeq5a_\mathrm{C}$, the latter magnified 6 times, marked by the dashed lines and arrows of the same colors in the field plot. (d)~As (c) after high-pass filtering above 5.3~MHz. (e)~Magnitude of the two-dimensional Fourier transform of (d), normalized to its maximum; points at right are the section at $f=16$~MHz, marked by the dashed line and arrow in the map, and the red line is the Lorentzian fit, with the numbers labeling the two modes. Color scales saturate at $\pm0.5$ in (a--b) and $\pm0.1$ in (c--d), in the same arbitrary units.}
  \label{fig:simulation}
\end{figure*}

We first consider transverse excitation in statistically isotropic polycrystalline iron, formed by cubic symmetry grains with elastic constants $c_{11}=219.2$, $c_{12}=136.8$ and $c_{44}=109.2$~GPa, density $\rho=7860$~kg/m$^3$, Zener anisotropy index $A=2.65$ and universal anisotropy index~\cite{Ranganathan2008} $A^U=1.23$. Figure~\ref{fig:simulation} illustrates the three-dimensional finite element simulation of a 16~MHz plane transverse wave pulse, for which the Voigt-averaged wavelength $\lambda_\mathrm{0T}=0.20$~mm is smaller than the average grain and $2k_\mathrm{0T}a_\mathrm{C}=14.37$. The computational domain, shown in Fig.~\ref{fig:simulation}(a), is a $12\times12\times5$~mm$^3$ cuboid containing 5760 equiaxed grains generated by Voronoi tessellation, with mean grain size $d=0.5$~mm and two-point correlation length $a_\mathrm{C}=0.23$~mm~\cite{Sha2020}. The domain is discretized into a uniform structured mesh of 369 million 0.0125~mm hexahedral elements, achieving a numerical error of about 0.1\%~\cite{Huang2020a}. The crystallographic orientations of the grains are assigned uniformly at random, creating a statistically isotropic aggregate. The grains are lossless, with no intrinsic dissipation, so all attenuation arises from scattering. A three-cycle, 16~MHz toneburst transverse force polarized along $x$ is applied uniformly over the $z=0$ face, and boundary conditions on the four lateral surfaces restrain their $y$ and $z$ motions to enforce plane-wave propagation along $z$.

The model is solved with the GPU-accelerated solver Pogo~\cite{Huthwaite2014}. Figure~\ref{fig:simulation}(a) shows the wavefield $u_x$ on the mid-$x$ plane at $t=0.7~\upmu$s, and Fig.~\ref{fig:simulation}(b) tracks its evolution. The wavefront is smooth at $t=0.16~\upmu$s, breaks up on the grain scale after roughly one correlation length at $t=0.34~\upmu$s, and converts fully to speckle at $t\geq0.70~\upmu$s. This means that the injected wave is extinguished at about one correlation length, and whatever coherence survives at depth must be carried by forward scattering. To extract that coherent component, the displacement is averaged over the $x$--$y$ cross section at each of 71 depths from $z=0$ to 3.5~mm in 0.05~mm steps, excluding planes near the back face at $z=5$~mm to avoid reflections, and then averaged over 20 independent realizations with re-randomized orientations, which is sufficient for convergence. The resulting transverse coherent field $u_x(t,z)$ is shown in Fig.~\ref{fig:simulation}(c); the $u_y$ and $u_z$ components are three orders of magnitude smaller and carry no coherent part. The coherent wave shows one streak near the source at $z=0$, while two streaks of different slopes appear beyond $z\simeq a_\mathrm{C}$, consistent with the extinction of the injected wave at one correlation length. Because attenuation grows with frequency, the two streaks are mostly masked by the surviving low-frequency components beyond $z\simeq5a_\mathrm{C}$, but emerge clearly after high-pass filtering in Fig.~\ref{fig:simulation}(d). Here and throughout, the filter is a zero-phase minimum-order IIR high pass with 60 dB stopband attenuation and passband edge $f_\mathrm{c}(N-2)/N$, the lower bandwidth limit of a $N$-cycle toneburst with center frequency $f_\mathrm{c}$. A two-dimensional Fourier transform~\cite{Alleyne1991} of $u_x(t,z)$ then maps the coherent field into the frequency--wavenumber plane in Fig.~\ref{fig:simulation}(e), where ridges of high amplitude trace the dispersion relations of the propagating coherent modes. Near 10~MHz, where $2k_\mathrm{0T}a_\mathrm{C}\simeq9$ and $\lambda_\mathrm{0T}$ falls below $\simeq1.4a_\mathrm{C}$, marking the onset of the geometric regime, two ridges emerge with distinctly different slopes and widths, indicating two coherent modes with their own phase velocities and attenuation coefficients.

The two modes are unambiguously resolved, constituting a direct observation of multiple coherent modes in a single-phase random medium.

For comparison we solve the Dyson equation for the mean Green's function in the first-order smoothing approximation, which yields the dispersion equation for the complex effective wavenumber $k(\omega)$~\cite{Karal1964,Frisch1968,Weaver1990,Sha2020}
\begin{equation}
    \omega^2-k^2V_{0M}^2-m_M(\mathbf{k},\omega)=0
\label{eq:dispersion}
\end{equation}
for excitation mode $M$, where $\omega=2\pi f$, $V_{0M}$ is the Voigt-averaged velocity of mode $M$, and $\mathbf{k}=k\mathbf{p}$ with $\mathbf{p}$ representing the propagation direction. Other reference media can be used~\cite{Kube2015}; we take the Voigt average, for which the theory agrees well with simulations~\cite{Huang2022a,Huang2022b}. The mass operator, or self-energy, $m_M$ sums the three scattering channels $M\rightarrow\mathrm{L}$, $M\rightarrow\mathrm{T1}$ and $M\rightarrow\mathrm{T2}$, where L denotes the longitudinal wave and T1 and T2 the two transverse waves. It is this channel structure that gives Eq.~\eqref{eq:dispersion} three roots~\cite{Huang2021}. Each root carries a phase velocity $V_M=\omega/\mathrm{Re}\,k$ and an attenuation coefficient $\alpha_M=\mathrm{Im}\,k$.

\begin{figure*}[!ht]
  \includegraphics[width=0.72\textwidth]{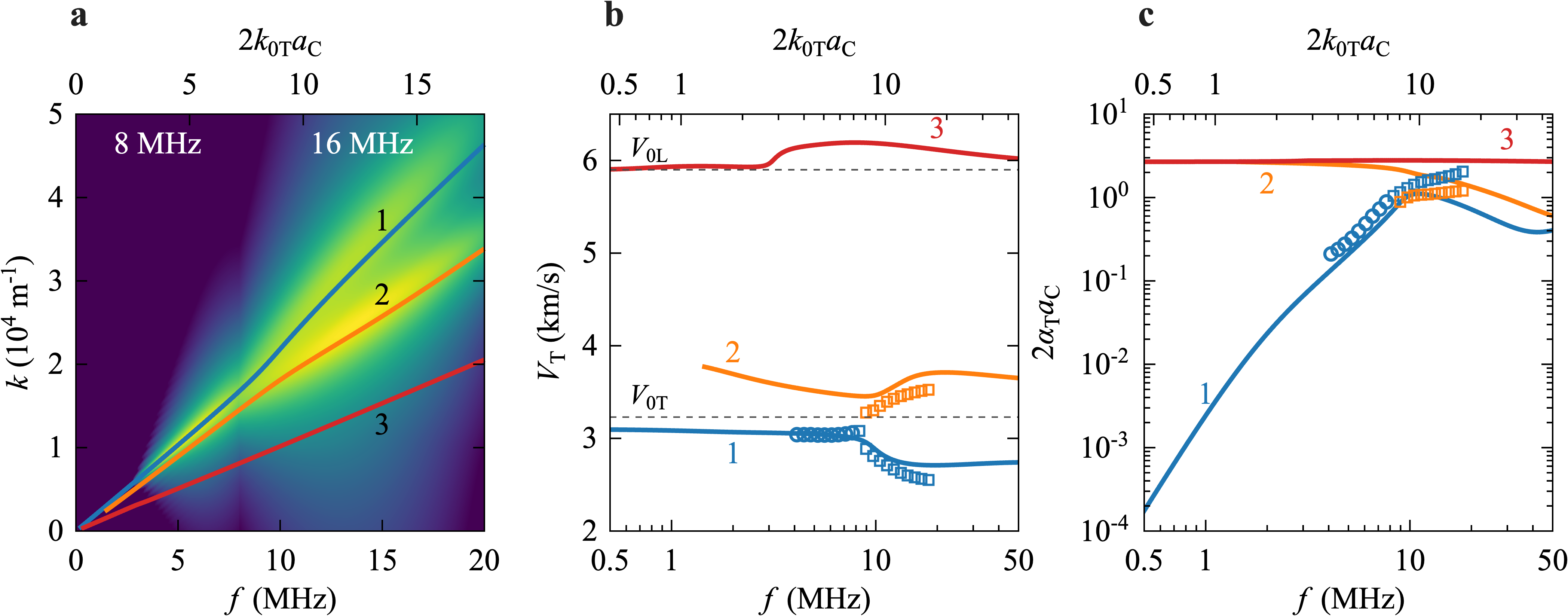}
  \centering
  \caption{Transverse wave propagation in iron: simulation against theory. (a)~Spectral magnitude of the high-pass-filtered coherent displacement from simulations with 8 and 16~MHz center frequency (labeled), each normalized to its own peak and shown on a logarithmic color scale; solid curves are the three roots of Eq.~\eqref{eq:dispersion}, numbered 1--3. (b)~Phase velocity $V_\mathrm{T}$ of the three roots; horizontal dashed lines mark the Voigt-averaged transverse ($V_\mathrm{0T}$) and longitudinal ($V_\mathrm{0L}$) velocities. (c)~Attenuation coefficient normalized with correlation length, $2\alpha_\mathrm{T} a_\mathrm{C}$. Symbols in (b) and (c) are the values extracted from the simulated spectra in (a) by Lorentzian fitting, colored as the corresponding root, one marker type per center frequency. Upper axes: normalized frequency $2k_\mathrm{0T}a_\mathrm{C}$, with $k_\mathrm{0T}=2\pi f/V_\mathrm{0T}$.}
  \label{fig:result_T}
\end{figure*}

The three roots for transverse propagation in iron are overlaid on the measured spectra in Fig.~\ref{fig:result_T}(a), which combines the 16~MHz simulation with an additional 8~MHz simulation on a longer, $12\times12\times10$~mm$^3$ domain, where the coherent signals are taken over $z=0$--4.5~mm and all else is unchanged. Root 1 follows the measured ridge closely across the entire frequency range and root 2 emerges above $2k_\mathrm{0T}a_\mathrm{C}\simeq9$ in the geometric regime. The third root has no counterpart in the simulation.

To compare quantitatively we extract each mode from the simulation spectra. Mode $i$ is a pole of the mean Green's function at complex wavenumber $k_i+j\alpha_i$ (with $j^2=-1$), so it propagates as $e^{jk_iz-\alpha_iz}$ and contributes to the power spectrum a Lorentzian centered on $k_i$ with half-width $\alpha_i$. Neglecting the interference between well-separated modes, we fit $\sum_{i=1}^{2}A_i/((k-k_i)^2+\alpha_i^2)+b$ at each frequency, as illustrated in Fig.~\ref{fig:simulation}(e), and obtain the measured points in Figs.~\ref{fig:result_T}(b) and \ref{fig:result_T}(c). Both modes are resolved in single realizations, whose cross-sectional average spans several hundred grains; between realizations their fitted velocities vary by less than 1\% and their attenuations by about 6\%.

Theory and simulation agree closely. The phase velocities of both modes are reproduced across the entire simulated frequency range, and the attenuation coefficients agree wherever a single mode propagates. In the geometric regime the measured attenuation deviates from the prediction. The deviation stems not from the Lorentzian fitting but from the theory, which is used here at $2\alpha a_\mathrm{C}\gtrsim1$, beyond weak scattering.

Theory and simulation explain the multiplicity. At low frequencies where $2k_\mathrm{0T}a_\mathrm{C}<9$, mode 1 propagates marginally slower than $V_\mathrm{0T}$, sampling the two transverse channels equally in an effective fashion, and its attenuation lies orders of magnitude below that of the other roots. This explains why conventional ultrasonic measurements observe one transverse wave, and only one. Beyond $2k_\mathrm{0T}a_\mathrm{C}\simeq9$ the geometric regime is reached, and propagation is better pictured as rays~\cite{Rokhlin1992} that travel between scattering events predominantly in one of the three channels of L, T1 and T2, with mode conversion linking them. The three roots inherit the three channel velocities, with mode 1 riding predominantly the slow transverse channel, mode 2 the fast one, and mode 3 the longitudinal channel. Crucially, in this regime modes 1 and 2 have attenuation coefficients of the same order, as seen in Fig.~\ref{fig:result_T}(c) from both theory and simulation, so both survive over comparable distances and both appear in the spectra. Mode 3 is different in kind. The repeated L--T conversion it requires, involving large-angle deflection and high-contrast transmission at grain boundaries, keeps its attenuation above the others across the entire frequency range, and this is the reason why it is never visible in our numerical experiments.

Longitudinal excitation completes the picture. The longitudinal wavelength is longer at a given frequency, and we therefore select SrMgSn, a cubic polycrystal of very strong anisotropy with $c_{11}=7$, $c_{12}=4$ and $c_{44}=25$~GPa, $\rho=3472.8$~kg/m$^3$, $A=16.67$ and $A^U=17.67$, so that the geometric regime is reached at computationally accessible frequencies~\cite{Huang2022b}. Simulations at 2, 4 and 8~MHz center frequency reuse the $12\times12\times10$~mm$^3$ domain for 2 and 4~MHz and the $12\times12\times5$~mm$^3$ domain for 8~MHz, with the polarization now along $z$ (namely force applied in $z$ direction, Fig.~\ref{fig:simulation}(a)) and symmetry boundary conditions on the lateral surfaces.

\begin{figure*}[!ht]
  \includegraphics[width=0.72\textwidth]{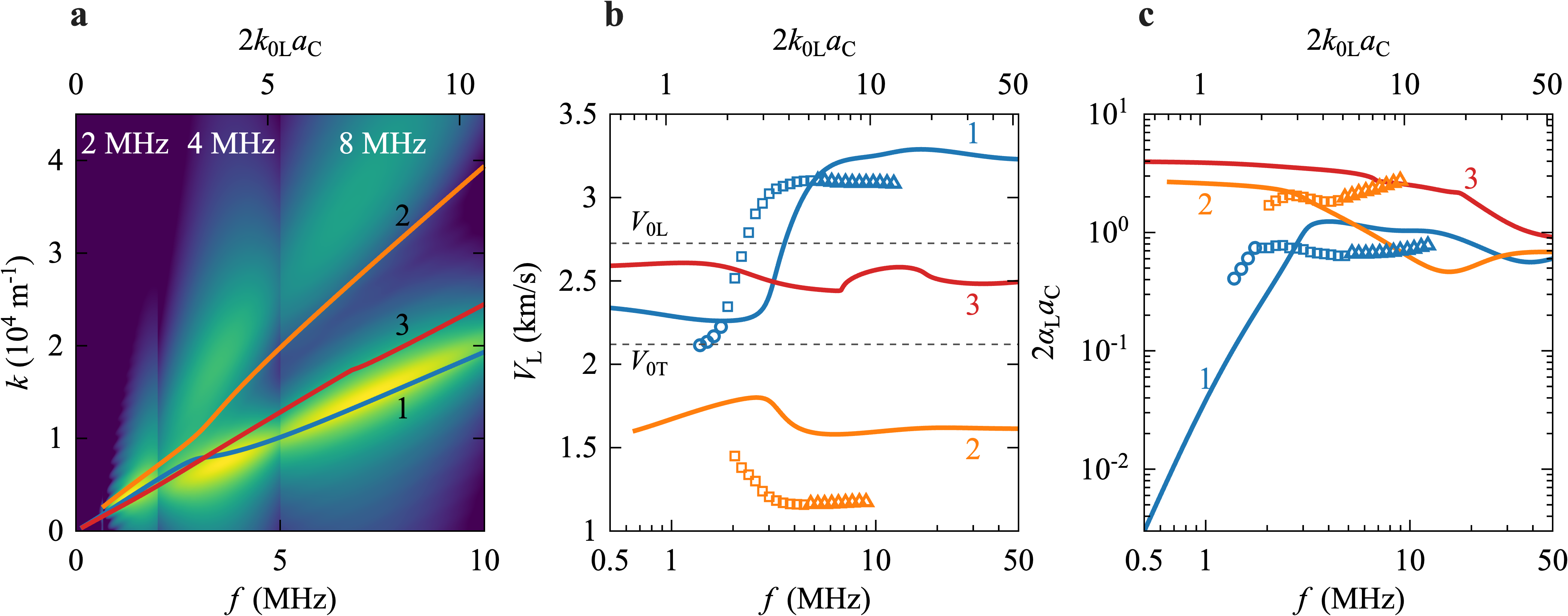}
  \centering
  \caption{Same as Fig.~\ref{fig:result_T}, but for longitudinal wave propagation in the strongly anisotropic polycrystal SrMgSn, simulated at 2, 4, and 8~MHz center frequency (labeled). Upper axes: normalized frequency $2k_\mathrm{0L}a_\mathrm{C}$, with $k_\mathrm{0L}=2\pi f/V_\mathrm{0L}$.}
  \label{fig:result_L}
\end{figure*}

Figure~\ref{fig:result_L}(a) shows the resulting spectra with the three theoretical roots overlaid, and the same phenomenon appears, with a single ridge at low frequencies and two ridges above about 3~MHz ($2k_\mathrm{0L}a_\mathrm{C}\simeq3$). The lower-$k$ ridge follows root 1 closely. The second ridge lies at higher $k$ and thus lower velocity than root 2 predicts, which is an expected discrepancy as the theory is known to overestimate phase velocity at this extreme anisotropy~\cite{Huang2022a}. In Figs.~\ref{fig:result_L}(b) and \ref{fig:result_L}(c) the quantitative agreement is looser than in iron, as the first-order smoothing approximation is pushed well outside the weak-scattering conditions for which it was derived. However, theory and simulation follow the same profile over the whole frequency range for both phase velocity and attenuation: they both show clear onset of the geometric regime. That an approximate theory retains this much at $A^U=17.67$, an anisotropy an order of magnitude beyond that of iron, indicates that the multimodal structure it predicts is robust and not an artifact of weak scattering.

These results repeat the transverse story with the roles rescaled by anisotropy. At low frequencies, mode 1 travels near $V_\mathrm{0L}$, with a larger offset than in iron commensurate with the higher anisotropy, and it is the only weakly attenuated root, so the effective-medium description holds and one longitudinal wave is seen. The transition now occurs at $2k_\mathrm{0L}a_\mathrm{C}\simeq3$ ($\lambda_\mathrm{0L}\simeq4a_\mathrm{C}$), earlier than in iron due to the stronger anisotropy. In the geometric regime, mode 1 speeds up toward the fast propagation paths, mode 2 is slowest with its propagation dominated by the slow transverse channel, and the two attain comparable attenuation, making both observable. Mode 3 propagates at an intermediate velocity but again attenuates faster than the others.

Two conclusions follow. First, the multiple roots of the effective dispersion equation are physical. An approximate theory built on the first-order smoothing approximation correctly predicts, for both polarizations and at both anisotropies, which modes appear, where in frequency they separate, and their dispersion. The extra roots are not artifacts of the approximation, and the agreement across the two very different materials indicates that the multimodal structure is a generic property of statistically isotropic polycrystals. The observation equally vindicates the causality argument of Calvet and Margerin~\cite{Calvet2012} that in the geometric regime, a single-mode description of the coherent field is not merely inaccurate but forbidden, and the modes that restore causality are in principle observable waves. Second, the number of observable modes is governed by the attenuation hierarchy of the roots. Equation~\eqref{eq:dispersion} has three roots at most frequencies. One mode is always observable, the second becomes observable in the geometric regime as the attenuation for both modes becomes comparable, while the third decays faster at all frequencies and is unobservable. This answers, for polycrystals, the question raised for particulate materials~\cite{Gower2019a}: of the many mathematically admissible effective waves, those that survive are the ones whose attenuation permits propagation over many microstructural lengths.

The mechanism identified here, in which distinct forward-scattering channels carry distinct coherent modes once the wavelength resolves the microstructure, invokes nothing specific to elasticity. The Dyson equation and its dispersion relation govern coherent propagation in all random media~\cite{Frisch1968,Sheng2006}, and comparable multimodality should arise for electromagnetic waves in random dielectrics and for high-frequency seismic waves in the heterogeneous Earth~\cite{Sato2012} and other planetary bodies; the result bears on theories of high-frequency energy transport~\cite{Weaver1990}.

In summary, grain-scale simulations resolve the coherent wavefield of statistically isotropic polycrystals into its constituent modes and reveal it to be multimodal in the geometric regime. The single effective elastic wave of the classical description holds through the Rayleigh and stochastic regimes, which cover most practical measurements, and it is in the geometric regime, reached here for the first time, that the remaining modes appear.

\bibliography{references}

\end{document}